\documentclass[draft]{agujournal2019}
\usepackage{url} 
\usepackage{lineno}
\usepackage{soul}
\usepackage{amsmath}
\usepackage{siunitx}
\usepackage[
    draft,
    commentmarkup=footnote
]{changes}

\definechangesauthor[name=Georg Sebastian, color=green]{GSV}

\drafttrue

\journalname{Journal of Advances in Modeling Earth Systems (JAMES)}

\begin{document}

\title{The transient IDEMIX model as a nonorographic gravity wave parameterization in an atmospheric circulation model}

\authors{B. Quinn\affil{1}, C. Eden\affil{1}, D. Olbers\affil{2,3}, G.S. Voelker\affil{4}, U. Achatz\affil{4}  }

\affiliation{1}{Institut f\"ur Meereskunde, Universit\"at Hamburg, Hamburg, Germany}
\affiliation{2}{Alfred Wegener Institute, Bremerhaven, Germany}
\affiliation{3}{Center for Marine Environmental Sciences, Bremen University, Bremen, Germany}
\affiliation{4}{Institut f\"ur Atmosph\"are und Umwelt, Goethe Universit\"at Frankfurt am Main, Frankfurt, Germany}

\correspondingauthor{Brenda Quinn}{brenda.quinn@uni-hamburg.de}

\begin{keypoints}
\item Transient 1D parameterization of internal gravity waves in the atmosphere
\item Zonal gravity wave drag reversal at mesopause
\item Alternate meridional wind structures
\end{keypoints}

\begin{abstract}
The Internal wave Dissipation, Energy and Mixing (IDEMIX) model presents a novel way of parameterizing internal gravity waves in the atmosphere.  Using a continuous full wave spectrum in the energy balance equation and integrating over all vertical wavenumbers and frequencies results in prognostic equations for the energy density of gravity waves in multiple azimuthal compartments. It includes their non-dissipative interaction with the mean flow, allowing for an evolving and local description of momentum flux and gravity wave drag. A saturation mechanism maintains the wave field within convective stability limits, and an energetically consistent closure for critical-layer effects controls how much wave flux propagates from the troposphere into the middle atmosphere.  IDEMIX can simulate zonal gravity wave drag around the mesopause, similar to a traditional gravity wave parameterization and to a state-of-the-art wave ray tracing model in an atmospheric circulation model.  In addition, IDEMIX shows a reversal of the gravity wave drag around the mesopause region due to changes in the momentum flux there.  When compared to empirical model data, IDEMIX captures well the summer hemisphere flow reversal, the cold summer mesospheric pole and the alternate positive and negative structures in the meridional mean flow.

\end{abstract}


 \section{Introduction}

Despite the improved computational grid resolutions of numerical weather prediction models, a recent study has shown that there is still a deficit of up to 30$\%$ of gravity wave forcing in the extratropical stratosphere at 9km horizontal resolution and 15$\%$ at 4km resolution~\cite{Polichtchouketal2023}.  It is generally accepted~\cite{Polichtchouketal2023,Kimetal2003}, that there is an ongoing need for gravity wave parameterizations (GWP) in the weather prediction and climate models in conjunction with the now partially-resolved gravity wave drag (GWD).

However, the level of sophistication of these GWP also requires enhancing.  One area where improvement is required is the level of physical processes included in sub-grid-scale parameterizations of numerical weather prediction and climate models~\cite{Kimetal2003}.  All current operational GWP are steady state, relying on wave breaking and the accompanying momentum deposition to induce GWD onto the mean flow. 
However, it was shown by the 
transient multiscale gravity wave model (MS-GWaM)~\cite{Bolonietal2016,Bolonietal2021} that the GWD imparted onto the mean flow by wave–mean-flow interaction can dominate the wave-breaking effect. They demonstrated that their
transient Wentzel–\-Kramer\-Brillouin (WKB) model with 
wave breaking disabled,
%
may produce, to leading order, a similar spatial distribution of wave energy and induced mean flow as large-eddy simulations 
of idealized cases representing the propagation of isolated gravity wave packets.

MS-GWaM is a Lagrangian ray-tracing model 
and computationally costly.
On the other hand, 
the complementary transient model IDEMIX~\cite{OlbersEden2013,EdenOlbers2014,OlbersEden2017A,EdenOlbers2017B} 
proposed here is based on the spectral energy balance of the wave field and has previously been successfully developed as a model for diapycnal diffusivity, induced by internal gravity wave breaking in the ocean.  
It has recently been applied to atmospheric gravity waves~\cite{Quinnetal2020}, where the integration of the energy balance equation for a continuous wave field of a given spectrum, results in prognostic equations for the energy density of eastward, westward, northward and southward gravity waves.  
IDEMIX is here implemented into the global Upper Atmosphere Icosahedral Nonhydrostatic (UA-ICON)~\cite{Borchertetal2019,Zangletal2015}.   MS-GWaM has also recently been implemented into UA-ICON~\cite{Bolonietal2021,Kimetal2021} which allows for a straightforward comparison between these two innovative transient gravity wave models.  It includes convective sources~\cite{Kimetal2021} and lateral propagation~\cite{Voelkeretal2023}, 
which have been shown to be important to the spatial distribution of GWD, especially during winter in the high latitudes of the southern hemisphere
and also to the structure of the quasi-biennial oscillation
~\cite{AmemiyaSato2016,Voelkeretal2023,Kimetal2023}.

The aim of this study is to 
to compare the performance of IDEMIX and MS-GWaM in UA-ICON 
and also  to 
the current nonorographic GWP employed by UA-ICON, which
is based on the work of~\citeA{WarnerMcIntyre1996}, which was simplified to a hydrostatic, non-rotational atmosphere by~\citeA{Scinocca2003,McLandressScinocca2005}.  It was first implemented into the European Centre for Medium-Range Weather Forecasts (ECMWF) model by~\citeA{Orretal2010} and is subsequently denoted by WMS.
UA-ICON is the upper atmosphere extension of the ICON general circulation model which extends the dynamical core from shallow to deep-atmosphere equations.  This amounts to taking into account the spherical shape of the atmosphere, gravitational field and the usually-discarded parts of the Coriolis force.  It also includes physics parameterizations which become relevant in the rarefied air, for example, molecular diffusion, ion drag and 
joule heating and chemical heating~\cite{Borchertetal2019}.

In Section 2 we will briefly outline the concept of the new GWP IDEMIX for the atmosphere. In Section 3, we give details about the model setup and the results in terms of simulated zonally averaged wave drag, zonal and meridional flow, and temperature.
We also discuss sensitivities with respect to forcing
amplitude and model parameters. The last section provides
a summary and discussion of the results.

\section{The IDEMIX model for the atmosphere}

A brief description of the derivation of the IDEMIX model is given here, full details can be found in~\citeA{Quinnetal2020} and references therein.
The model concept is based on calculating 
bulk propagation and other
parameter of the closure based on an assumed spectral shape.
The Desaubies spectrum \cite{Desaubies1976,VanZandtFritts1989,FrittsVanZandt1993,FrittsLu1993,WarnerMcIntyre1996} is employed here in the form
\begin{eqnarray}
\label{Desaubies_spectrum}
\mathcal{E}(z,t,m,\omega,\phi) = E(z,t,\phi)\,A(m)\,B(\omega)\,,
\end{eqnarray}
where $A(m)= \hat{A}(\mu)/m_*$, $\mu=|m|/m_*$.  The wavenumber dependent part $\hat{A}(\mu)$ has the same shape at each frequency,
\begin{eqnarray}
\label{Amu}
\hat{A}(\mu) =  A_0\frac{\mu^q}{1+\mu^{q+r}} \quad {\rm{where}} \quad  A_0 = [(q+r) / \pi]\sin{[\pi(q+1)/(q+r)]}
\end{eqnarray}
and $m_{*}$ characterises the spectral peak and also its bandwidth in vertical wavenumber space.  The frequency dependent part is given by
\begin{eqnarray}
\label{Bomega}
 B(\omega) = B_0 \omega^{-p} \quad {\rm{where}} \quad B_0 = (p-1)f^{p-1}[1-(f/N)^{p-1}]^{-1}, \, p \neq 1.
 \end{eqnarray}
The normalisation factors $A_0$ and $B_0$ ensure that $\int_0^{\infty}\hat{A}(\mu) d\mu = 1$ and $\int^N_f B(\omega) d\omega = 1$.  
The nominal values of $(p,q,r) = (5/3,1,3)$ are chosen~\cite{Quinnetal2020,WarnerMcIntyre1996} since the coefficients vary only by a factor of two for plausible departures from these values~\cite{FrittsVanZandt1993}.  Note that the energy amplitude $E(z,t,\phi)$ is a density in wave direction $\phi$ and upon integrating over appropriate direction ranges, normalization must also be taken care of. 

This spectrum in Eq.~\eqref{Desaubies_spectrum} is inserted into the energy balance 
\begin{eqnarray}
\label{RTE1D}
\partial_t \mathcal{E} + \partial_z( \dot{z} \mathcal{E})  + \partial_m (\dot{m} \mathcal{E} ) + \partial_{\omega}(\dot{\omega}\mathcal{E})  = \dot{\omega}\mathcal{E}/\omega +\omega S\,,
\end{eqnarray}
which is equivalent to wave action conservation~\cite{BrethertonGarrett1968} for waves in the presence of a vertically sheared background flow ${\bf U}(z,t)$. 


The terms on the left-hand side of Eq.~\eqref{RTE1D} are the time variation of the wave field amplitude and the corresponding group velocities in $z$, $m$ and $\omega$ spaces respectively.  The first term on the right-hand side is the interaction term between the wave field and the mean flow.  The $S$ term denotes sources and sinks
and other non-linear effects.
The single column approach is assumed so that horizontal derivatives of the wave field do not appear.   
Eq.~\eqref{RTE1D} is also the basis of 
the columnar implementation of 
MS-GWaM, but formulated for
wave action $\mathcal{E}/\omega$. The difference between IDEMIX
and MS-GWaM is that for the  former we will integrate in phase
space, while the latter resolves those dimensions.

In the presence of a vertically-varying horizontal mean flow ${\bf U}(z,t)$, the frequency of encounter $\omega_{enc}$ of a wave is Doppler-shifted relative to the mean flow to the dispersion relation $\omega$ by
\begin{equation}
\label{Dopplershift}
\omega_{enc}({\bf k},m,z,t) =\omega + {\bf k}\cdot {\bf U}= \Omega({\bf k},m,z)+{\bf k}\cdot {\bf U}(z,t) = \Omega_{enc}({\bf k},m,z,t)\,
\end{equation}
where the local Boussinesq intrinsic dispersion relation for internal GWs~\cite{Sutherland2010} is defined as
\begin{eqnarray}
\label{GWdispersion}
\omega=\Omega({\bf{k}},m,z) =\left(\frac{N^2k^2+f^2m^2}{k^2+m^2}\right)^{1/2}\,.
\end{eqnarray}
The Doppler-shifted frequency in Eq.~\eqref{Dopplershift} is used to define the derivatives in Eq.~\eqref{RTE1D}, namely the vertical group velocity $\dot{z} =\partial\Omega/\partial m$ and the corresponding transport velocity in vertical wavenumber space by wave refraction 
$\dot{m} = -\partial_z\Omega_{enc}$.
The change in frequency is 
$\dot{\omega}= \dot{\omega}_{enc}-{\bf k}\cdot \dot{{\bf U}} = {\bf k}\cdot \partial_t {\bf U} -{\bf k}\cdot ( \partial_t {\bf U} +\dot{z}  \partial_z {\bf U}) = -{\bf k}\cdot \dot{z}  \partial_z {\bf U}$.




With these substitutions, Eq.~\eqref{RTE1D} is then integrated over all possible negative wavenumbers, for upward propagation, $m \in (-\infty,0)$  and frequencies $\omega \in (f,N)$.  

When the phase speed of a wave equals the mean flow velocity, the altitude at which this occurs is defined as a critical level (CL).  These are accounted for in IDEMIX in an energetically-consistent manner by means of the wave flux term in the vertical wavenumber $m$-space.  The theoretical behaviour of the waves at a CL is such that the vertical wavenumber goes to infinity while in practical terms, this wavenumber will be very large, rather than infinite.  This effect can be included in IDEMIX via a flux across a high wavenumber cut-off $m_s$ in wavenumber space.  Integration of $\partial_m (\dot{m} \mathcal{E} )$ over frequency, direction range and vertical wavenumber yields
\begin{eqnarray}
\label{2ndbouflux}
\int_f^N \dot m \, \mathcal{E}_i    d\omega\Big|^0_{-m_s}
& = & - A(-m_s)E_i\int^{N}_f\dot{m}(-m_s)B(\omega) d\omega\,,
\end{eqnarray}
where $i$ denotes the relevant azimuthal energy compartment, east, west, north, south, defined below.
The change of the vertical wavenumber along the ray is
\begin{eqnarray}
\label{mdot}
\nonumber
\dot{m} &=& - \partial_z \Omega - k C \\
 &=& - \frac{N}{\omega}\frac{\omega^2-f^2}{N^2-f^2}\frac{dN}{dz}- |m| \sqrt{\frac{\omega^2-f^2}{N^2-f^2}}C \,,
\end{eqnarray}
where $C = U_z\cos\phi  + V_z\sin\phi $.
The first term on the right-hand side is neglected for now as it only generates turning points but will be included in a future version of IDEMIX.  The energy flux defined in Eq.~\eqref{2ndbouflux} at the high wavenumber $-m_s$ then becomes
\begin{eqnarray}\label{alphacl}
m_s A(-m_s) \int^{N}_f \sqrt{\frac{\omega^2-f^2}{N^2-\omega^2}}  B(\omega) d\omega \;  E_i C  = \alpha \;  E_i C 
\end{eqnarray}
For the use as a CL parameterisation the flux in Eq.~\eqref{alphacl} is restricted 
 to be positive. 
Thus far, the IDEMIX model derivation from the energy balance equation has been energetically consistent.  

To prevent the wave amplitudes becoming too large, a Lindzen-type~\cite{Lindzen1981} saturation is employed, similar to many existing GWP.  The saturation coefficient is stated here with full derivation details found in~\citeA{Quinnetal2020,Bolonietal2016}.  The convective stability condition leads to a damping coefficient,
\begin{eqnarray}
\label{gamma}
\gamma = K \int^0_{-m_t} m^2 A(m) dm = \frac{K m_*^2}{\pi} \left(\ln\left(\frac{m_t}{m_*}\right)^4 + 1 \right)\,,
\end{eqnarray}
where 
\begin{eqnarray}
\label{TEdiffusiv}
K(z)=K_0\Big|\frac{m_*^2}{\pi \rho_0 N^2}\left(\ln\left(\frac{m_t}{m_*}\right)^4 + 1 \right)\,E_i  - \delta \Big|
\end{eqnarray}
derives from the stability condition.  It is set to zero for stable conditions and increases under unstable conditions. $K_0$ is a free parameter to control the magnitude of the damping coefficient, $\delta \le 1$ represents a parameter accounting for uncertainties of the criterion as in \cite{Bolonietal2016} and $m_t$ is the wavenumber at which turbulence sets in.  Based on estimates in the literature~\cite{Hines1991}, a turbulence cutoff wavenumber of $m_t=2\pi/100\,\si{m^{-1}}$ is adopted.

After the integration of Eq.~\eqref{RTE1D}
in $m$ and $\omega$,  integrating also
in four directional compartments for wave direction $\phi$,
and using the closures discussed above,
the IDEMIX model is given as
\begin{eqnarray}
\label{IDEMIX}
\nonumber
\partial_t E_e + \partial_z( c_0 E_e)  &=&  -\frac{\sqrt{8}}{\pi} \Lambda \partial_z U E_e  - \gamma_e E_e  -\alpha E_e \;\partial_z U   \\ 
\nonumber \partial_t E_w + \partial_z( c_0 E_w)
&=&  +\frac{\sqrt{8}}{\pi}  \Lambda \partial_z U E_w - \gamma_w E_w  -\alpha E_w\; \partial_z U  \\
\nonumber \partial_t E_n + \partial_z( c_0 E_n)  &=&  -\frac{\sqrt{8}}{\pi} \Lambda \partial_z V E_n - \gamma_n E_n  -\alpha E_n \;\partial_z V \\ 
\partial_t E_s + \partial_z( c_0 E_s)
&=&  +\frac{\sqrt{8}}{\pi}  \Lambda \partial_z V E_s  - \gamma_s E_s  -\alpha E_s\; \partial_z V\,.  
\end{eqnarray}
The azimuthal ranges for integration of the energy compartments $E_e$, $E_n$,
$E_w$, and $E_s$ are $(-\pi/4,\pi/4)$, $(\pi/4,3\pi/4)$, $(3\pi/4,5\pi/4)$, $(5\pi/4,7\pi/4)$, for east, north, west and south, respectively.  The mean group velocity $c_0$ and the mean flow interaction term $\Lambda$ result from the integration of the $\dot{z}$ and $\dot{\omega}/\omega$ terms from Eq.~\eqref{RTE1D} with their full expressions given in~\ref{apx_coeffs}.  
The basic setup of IDEMIX for the atmosphere is given by
Eq.~\ref{IDEMIX} and with a forcing condition for the vertical energy density flux (J $m^{-2}$ $s^{-1}$), $F_{bot} = c_0 E_{i}$ in each compartment, just above the tropopause.

\section{Results: IDEMIX in the UA-ICON model} 

\subsection{Experimental setup}
IDEMIX is implemented into the ICON-a model by means of a new module to the main ICON code.  It is validated against the current standard WMS nonorographic parameterization and also the cutting-edge transient MS-GWaM model, using the same columnar 
setup as~\citeA{Bolonietal2021}.  For MS-GWaM results presented here, a slightly older version of UA-ICON was used compared to the IDEMIX and WMS simulations.  A grid with horizontal spacing of approximately 160km (R2B04), 120 vertical layers to a grid height of 150km is employed. The ICON simulations are initialized with operational Integrated Forecasting System (IFS) of the ECMWF analyses for 1st May or 1st November 1991.  The data is extrapolated to the model top.  Two month simulations are performed with results showing the field averaged over the month of June or December, thus discarding the time for the mean field to settle at higher altitudes after initialization.
For all simulations, the standard orographic parameterization in ICON-a is always switched on and `no GWP' results shown refer only to no non-orographic GWP.

IDEMIX results are also compared to observational data from the UARS (Upper Atmosphere Research Satellite) Reference Atmosphere Project (URAP) and the Horizontal Wind Model (HWM14).
The UARS (Upper Atmosphere Research Satellite) Reference Atmosphere Project (URAP) describes a comprehensive reference atmosphere based on the UARS~\cite{SwinbankOrtland2003}.  A dataset provides monthly-mean zonal-mean zonal winds (not meridional) from the surface to the upper mesosphere for each month from January 1991 to December 1998.  Wind measurements from the High Resolution Doppler Imager (HRDI) were combined with results from The Met. Office stratospheric data assimilation system. Balanced winds derived from the URAP temperature data set were used to bridge the gap between the stratospheric winds and HRDI mesospheric winds.
The updated Horizontal Wind Model (HWM14) is another empirical reference atmosphere which provides zonal and meridional wind profiles from the surface to the lower thermosphere~\cite{Drobetal2015}.  Based on satellite observations and ground-based measurements, it provides a time-dependent global reference atmosphere.

A simple forcing, or boundary condition, for IDEMIX with a latitudinal variation is employed whereby the forcing has a slightly larger magnitude in the winter hemisphere than in the summer hemisphere.  This is to account for the seasonal variability of nonorographic GW sources emitted by jets and fronts, as done also in the MS-GWaM~\cite{Bolonietal2021} model case for comparison.  However for IDEMIX, it has been determined that by applying a forcing flux equivalent to the customary 2.5mPa launch momentum flux at $p=300\unit{mbar}$,
as adopted by WMS and MS-GWaM, gives results which are too unrealistic in the middle atmosphere.  It is thus adopted for IDEMIX, a basic standard forcing of 0.25mPa at $p=100\unit{mbar}$ and some results are also shown with 0.025mPa forcing magnitude.


\subsection{Basic IDEMIX results}


\begin{figure}[h!]
\includegraphics[width =0.5\columnwidth]{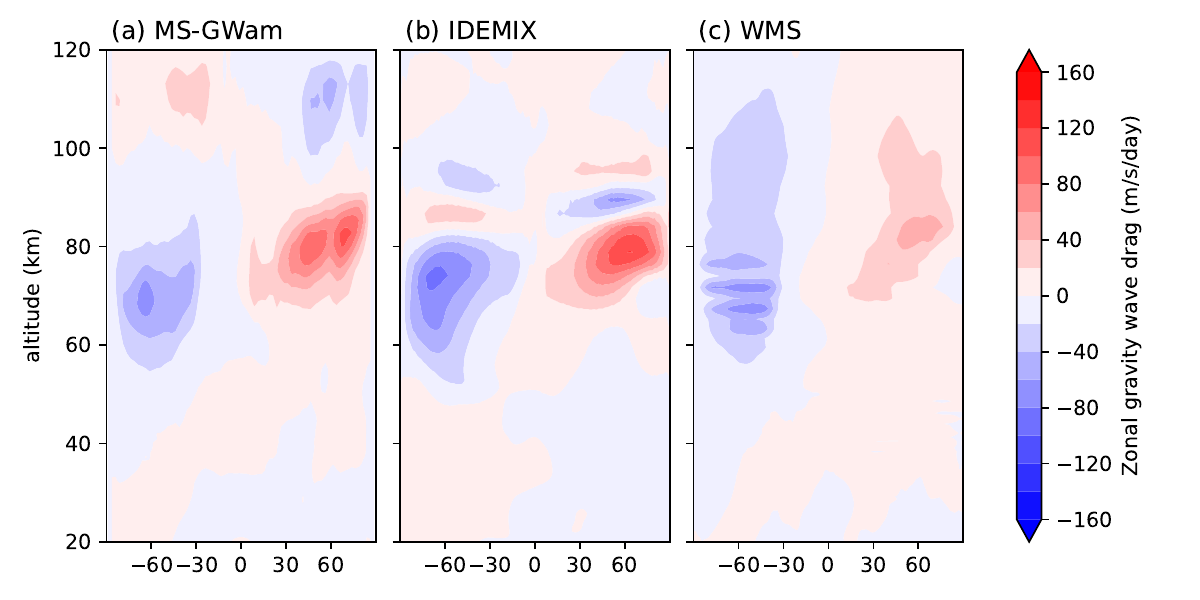}\includegraphics[width =0.5\columnwidth]{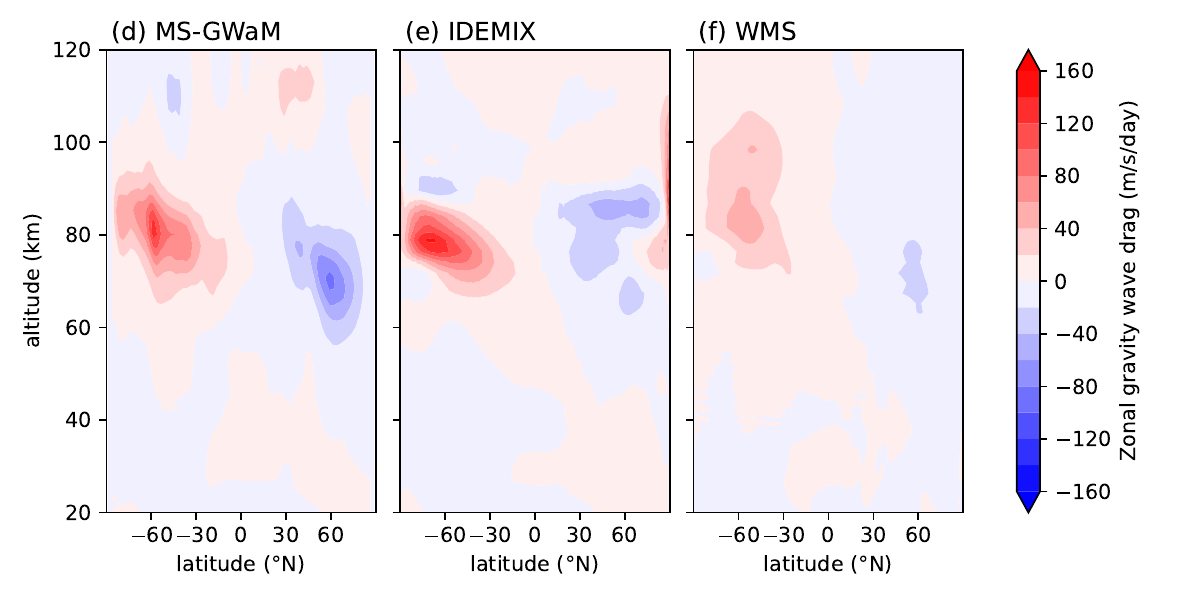}
\caption{ Zonal GWD (m/s/day) in a) to c) June 1991 and d) to f) December 1991. Forcing flux $F_{bot} = 2\times 10^{-2}$Jm$^{-2}$s$^{-1}$, $K_0 = 20$m$^2$s$^{-1}$, $m_s = 4.5\times 10^{-4}$m$^{-1}$ and $m_* =2\pi/15$km.}
\label{GWDU_jun_dec}
\end{figure}

Fig.~\ref{GWDU_jun_dec} shows the zonally-averaged zonal GWD 
for the different GW closures.
IDEMIX gives in general 
similar GWD distributions to MS-GWaM and WMS, but they all differ
in detail.
The positive and negative maxima for all three GWPs are comparable magnitudes and at comparable altitudes in both June and December.  In June, one notable difference with the IDEMIX distribution in Fig.~\ref{GWDU_jun_dec}b, is the thin band of GWD of the opposite sign at higher latitudes, just above the main maxima i.e. at about 85km altitude.  This change in GWD sign is also observed in December (Fig.~\ref{GWDU_jun_dec}e) but is much weaker.  This is not observed at all with WMS whereas for MS-GWaM, a weak change in sign of GWD occurs at around 100km altitude in both June and December.

\begin{figure}[h!]
\centerline{\includegraphics[width =0.8\columnwidth]
{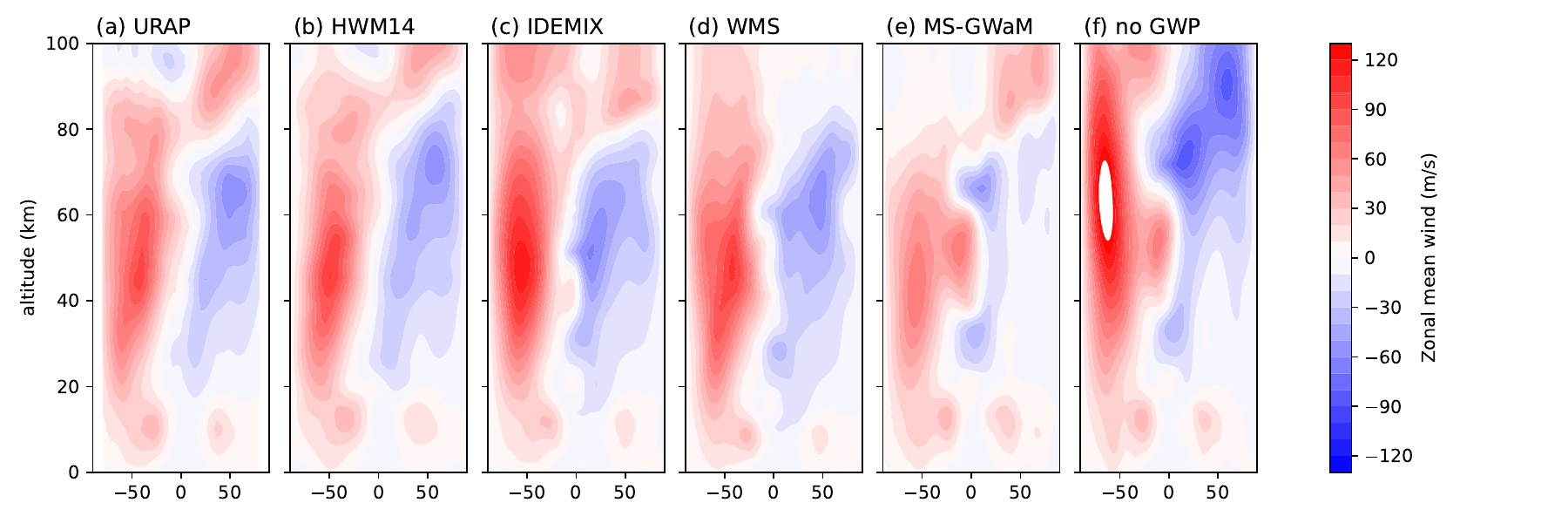}}

\centerline{\includegraphics[width =0.8\columnwidth]
{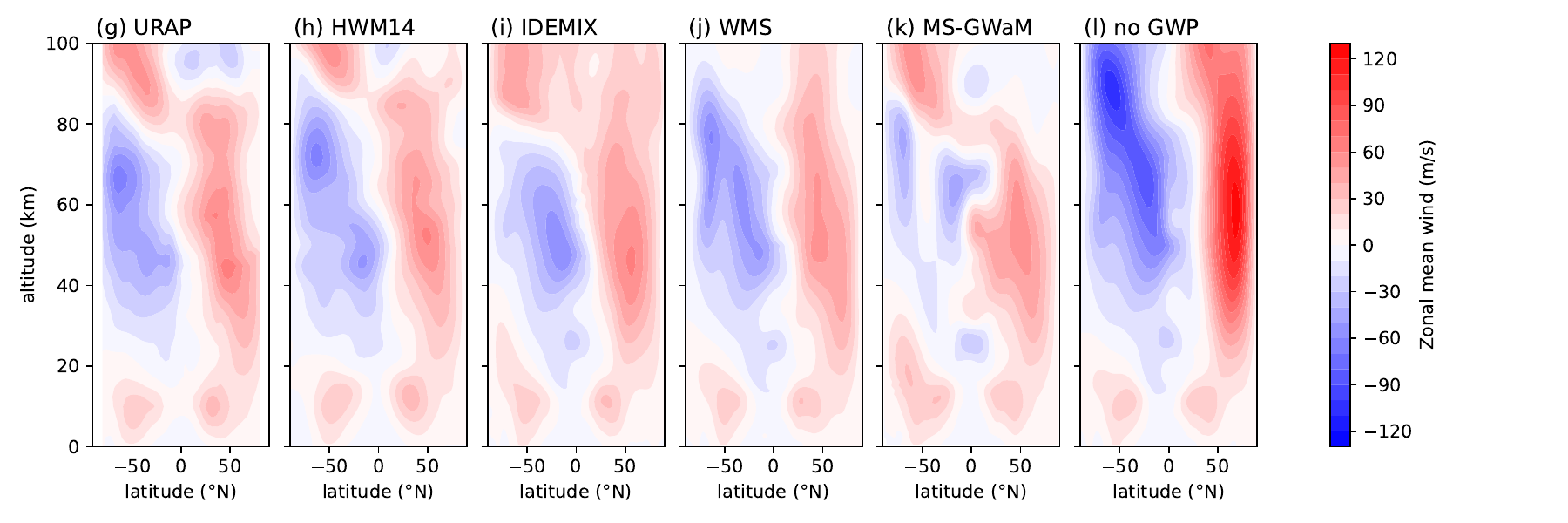}}

\caption{ Zonal mean flow velocity (m/s) in top June, bottom December 1991. Forcing flux $F_{bot} = 2\times 10^{-2}$Jm$^{-2}$s$^{-1}$, $K_0 = 20$m$^2$s$^{-1}$, $m_s = 4.5\times 10^{-4}$m$^{-1}$ and $m_* =2\pi/15$km. }
\label{U_jun_dec}
\end{figure}

\begin{figure}[h!]

\centerline{\includegraphics[width =0.8\columnwidth]{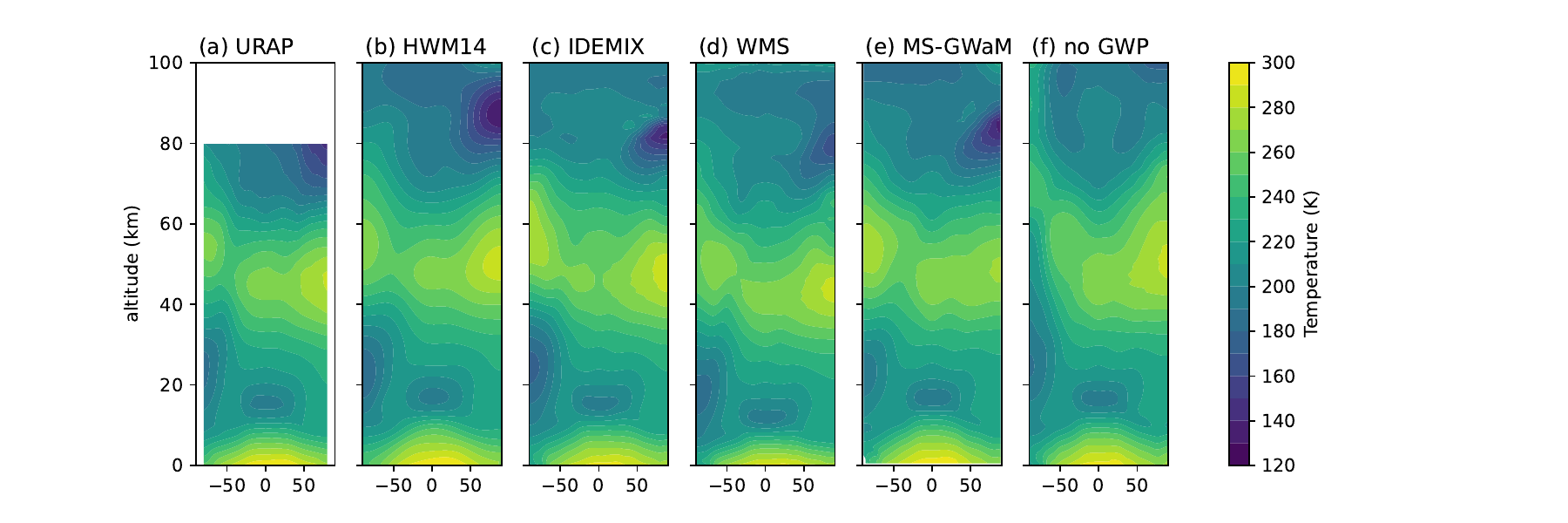}}

\centerline{\includegraphics[width =0.8\columnwidth]{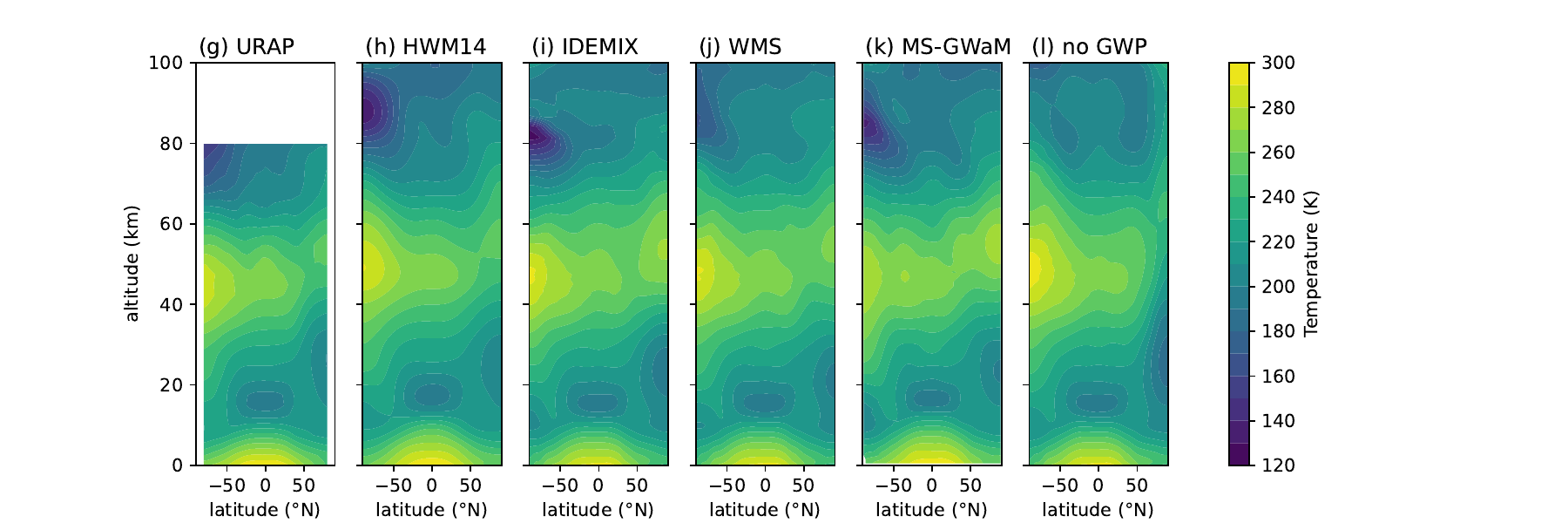}}
 
\caption{ Temperature (K) in top June, bottom December 1991. Forcing flux $F_{bot} = 2\times 10^{-2}$Jm$^{-2}$s$^{-1}$, $K_0 = 20$m$^2$s$^{-1}$, $m_s = 4.5\times 10^{-4}$m$^{-1}$ and $m_* =2\pi/15$km.}
\label{temp_jun_dec}
\end{figure}

\begin{figure}
\centerline{\includegraphics[width =0.8\columnwidth]{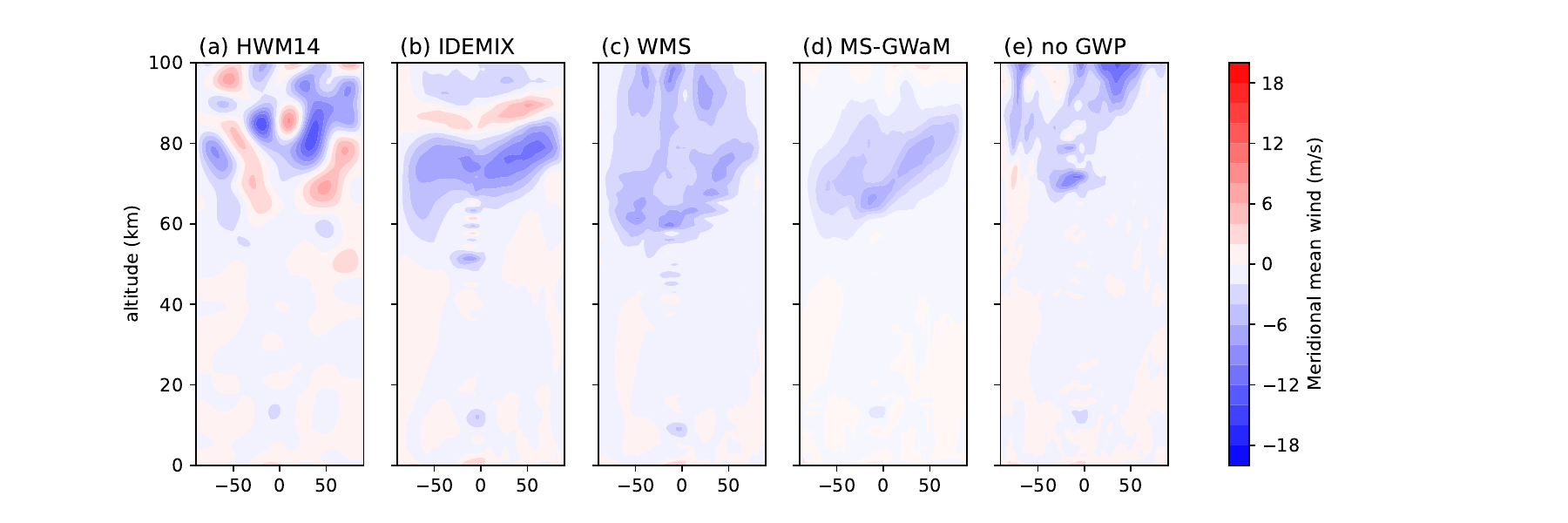}}

\centerline{\includegraphics[width =0.8\columnwidth]{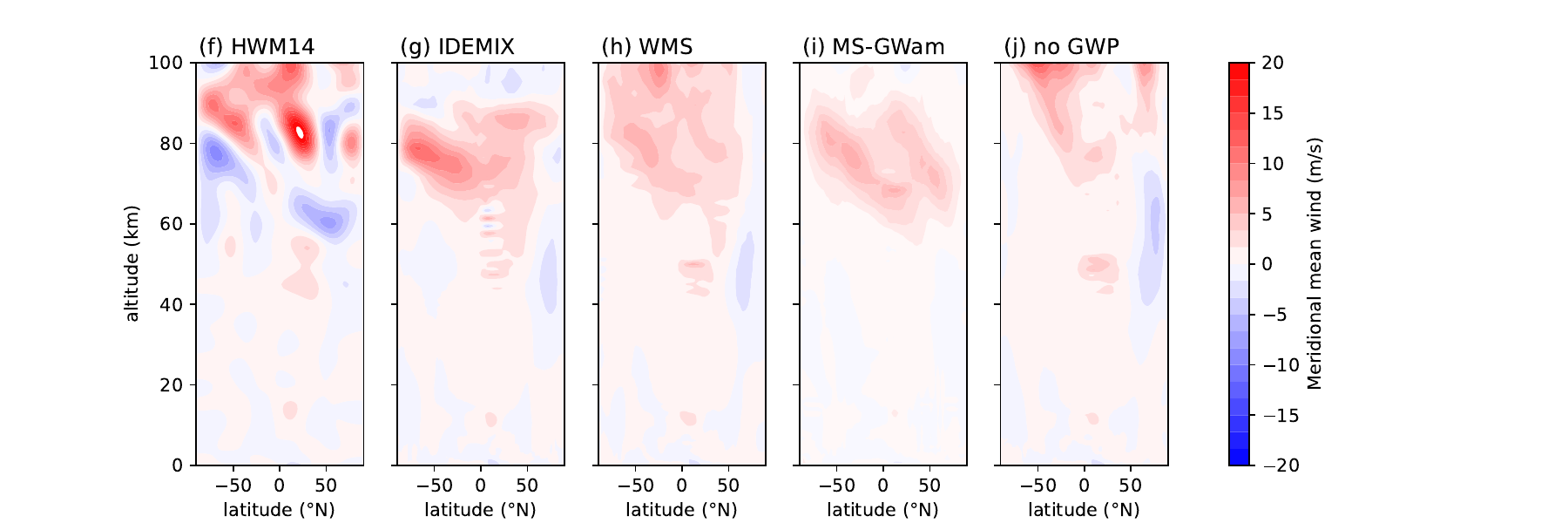}}

\caption{ Meridional mean flow velocity (m/s) for top: June 1991 and bottom December 1991. Forcing flux $F_{bot} = 2\times 10^{-2}$Jm$^{-2}$s$^{-1}$, $K_0 = 20$m$^2$s$^{-1}$, $m_s = 4.5\times 10^{-4}$m$^{-1}$ and $m_* =2\pi/15$km.}
\label{V_jun_dec}
\end{figure}

Fig.~\ref{U_jun_dec} compares the performance of the different
GWPs in UA-ICON with each other and the observational estimates
with respect
to the zonally-averaged zonal mean wind. Also the model
simulation without any GWP is shown.  Figs.~\ref{U_jun_dec}(f) and (l) show that without any GWP, there are large biases in the summer stratosphere and mesosphere, there is no flow reversal at the summer mesopause and the winter jet is much too strong.  Without a GWP, there is no cold summer mesopause or warm summer stratopause as evident in Figs.~\ref{temp_jun_dec}(f) and (l), and the zonally-averaged meridional flow is almost exclusively negative in June and positive in December, as shown in Figs.~\ref{V_jun_dec}(e) and (j) above 60km, contrary to observations.

In principle, all GWPs reduce the large biases in the simulation without GWP, including IDEMIX.
Figs.~\ref{U_jun_dec}(c) and (i) show that IDEMIX compares quite well to the URAP and HWM14 data in Figs.~\ref{U_jun_dec}(a) and (b) respectively in June and Figs.~\ref{U_jun_dec}(g) and (h) for December.  
IDEMIX exhibits the mean flow reversal in the summer hemisphere.  However this reversal for IDEMIX, as is the case for MS-GWaM in Fig.~\ref{U_jun_dec}(e), is at a lower altitude than the HWM14 data.  HWM14 data in Fig.~\ref{U_jun_dec}(b) show that this reversal should be at around 75km equatorward and around 90km polewards.  By contrast, IDEMIX and MS-GWaM have the reversal a few kilometres lower at 70km equatorward and 80km poleward. 
The WMS scheme does not show the mean flow reversal as such but the winter jet is decelerated to zero at 60km altitude equatorward and 85km polewards.  These altitudes correspond with the positive GWD maxima presented in Fig.~\ref{GWDU_jun_dec}.

The winter jet for IDEMIX is at a reasonable latitude and altitude and magnitude when compared to HWM14 data in Figs.~\ref{U_jun_dec}(b) and (h).  The summer jet for IDEMIX in December coincides with the HWM14 data lower maxima but IDEMIX does not exhibit a second at higher latitudes and altitude.  In June, as shown in Fig.~\ref{U_jun_dec}(c), the IDEMIX summer jet is also at a lower latitude and altitude than data suggests.
Also notable in the zonal mean flow distribution for December, is the IDEMIX summer tropospheric jet is vertically-elongated, as is the case for MS-GWaM (Figs.~\ref{U_jun_dec}(i) and (k) respectively).  This is not evident in the HWM14 data and for IDEMIX this is due to the forcing being too strong.

The aforementioned distributions of positive GWD also correspond to the location of the cold summer pole, as shown in Fig.~\ref{temp_jun_dec}.  Figs.~\ref{temp_jun_dec}(f) and (l) show that without a GWP, ICON does not exhibit a cold pole above the mesopause in neither June nor December.  IDEMIX simulates a cold pole of about 120K being centred just above the GWD maxima.  This is the case in both June and December, occurring at 82km for IDEMIX and 85km for MS-GWaM.  According to HWM14 data, this should be closer to 90km altitude.  In both June and December, the WMS scheme exhibits a warmer than reality pole of around 180K but the occurrence also coincides with the positive GWD distribution shown in Fig.~\ref{GWDU_jun_dec}.

\begin{figure}
\includegraphics[width =0.5\columnwidth]{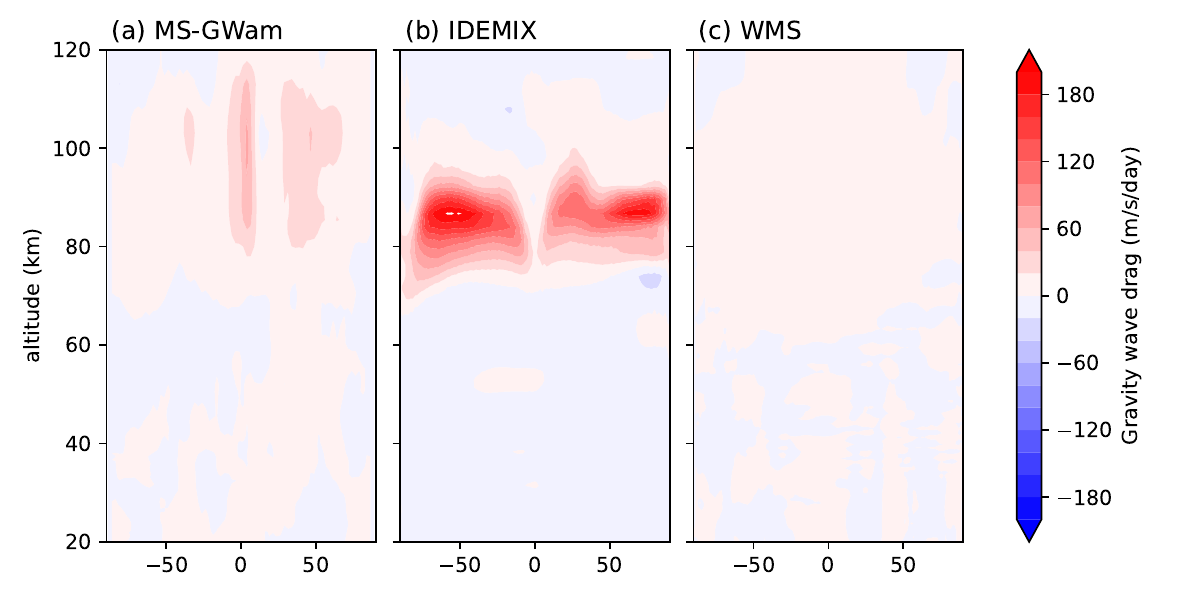}
\includegraphics[width =0.5\columnwidth]{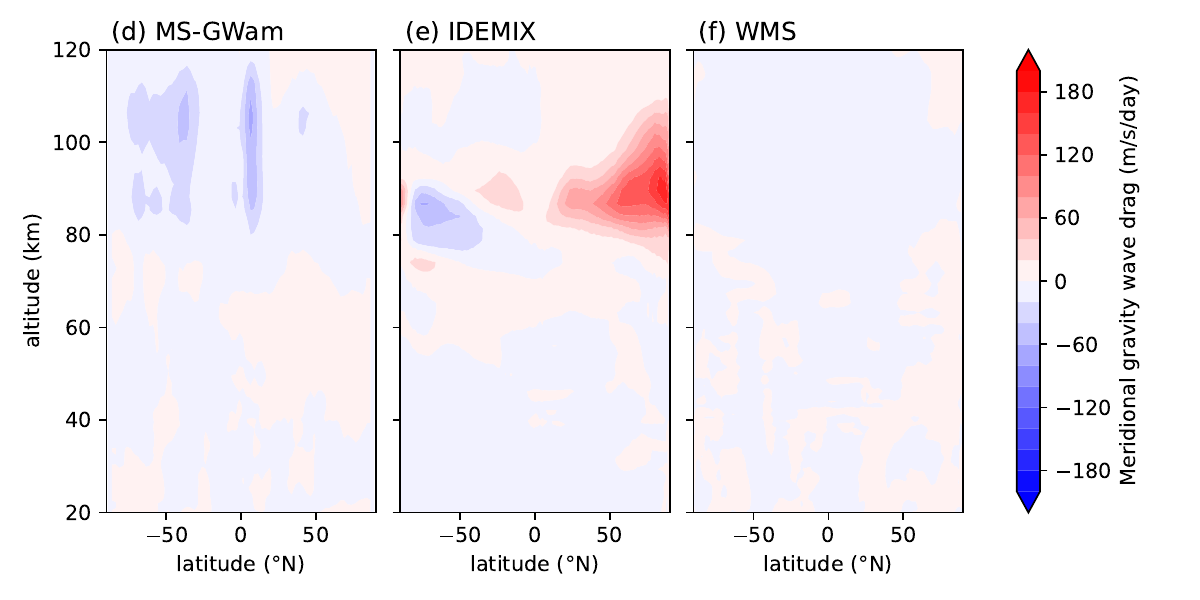}
\caption{ Meridional GWD (m/s/day) in (a)-(c) June 1991 and (d)-(f) December 1991. Forcing flux $F_{bot} = 2\times 10^{-2}$Jm$^{-2}$s$^{-1}$, $K_0 = 20$m$^2$s$^{-1}$, $m_s = 4.5\times 10^{-4}$m$^{-1}$ and $m_* =2\pi/15$km.}
\label{GWDV_jun_dec}
\end{figure}

The most profound difference which IDEMIX presents from the other GWPs is in the corresponding meridional mean flow and meridional GWD distributions which are shown in Figs.~\ref{V_jun_dec} and~\ref{GWDV_jun_dec} for the June and December 1991 cases presented so far.  Figs.~\ref{V_jun_dec}(b) and (g) show that IDEMIX exhibits more altitude-dependent structures to the meridional flow, as also present in the HWM14 data but it cannot be claimed that the structures of IDEMIX match those of HWM14.  
The IDEMIX structures are more vertically layered. Both WMS and MS-GWaM give purely negative meridional mean flow above the stratopause in June and only positive in December due to their very weak levels of simulated meridional GWD, as shown in Figs.~\ref{GWDV_jun_dec}(a) and (c). 
In June, IDEMIX shows also mainly positve GWD in both hemispheres, similar to MS-GWaM and WMS, although IDEMIX has a much greater magnitude of meridional GWD.  Conversely, in December, whereas MS-GWaM and WMS show mainly negative GWD, IDEMIX exhibits a positive GWD in the northern hemisphere and also at low southern latitudes in Fig.~\ref{GWDV_jun_dec}(e).

\subsection{Effects of wave forcing}



\begin{figure}[h!]
\centerline{\includegraphics[width = 0.8\columnwidth]{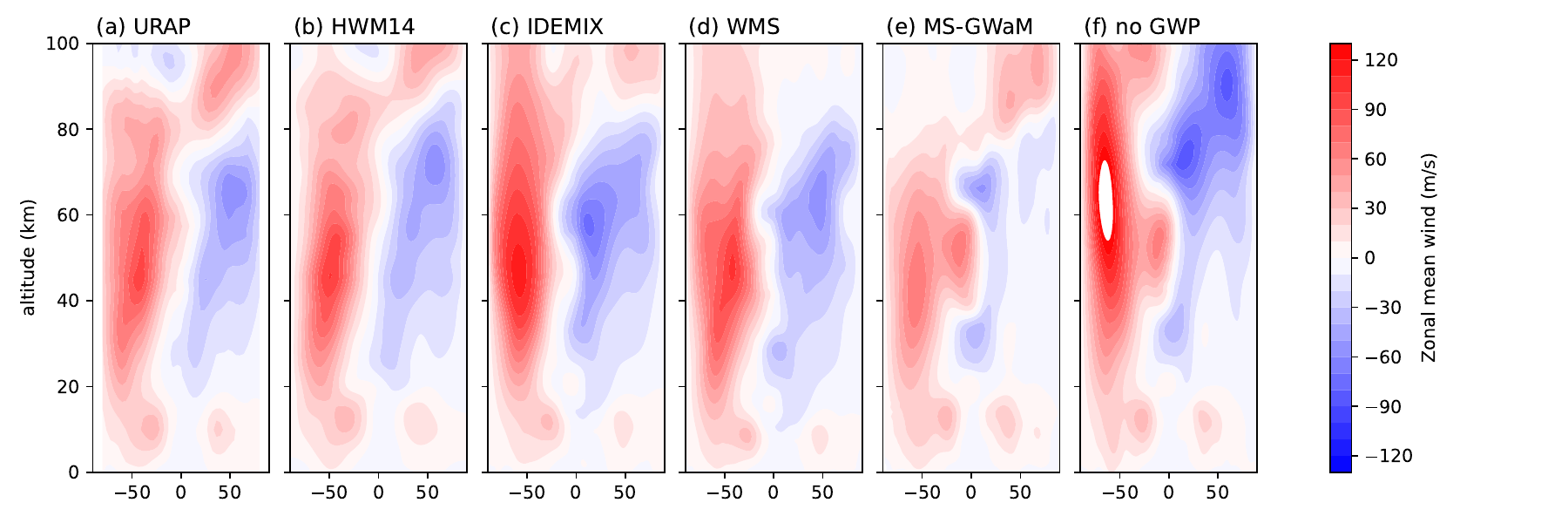}}

\centerline{\includegraphics[width = 0.8\columnwidth]{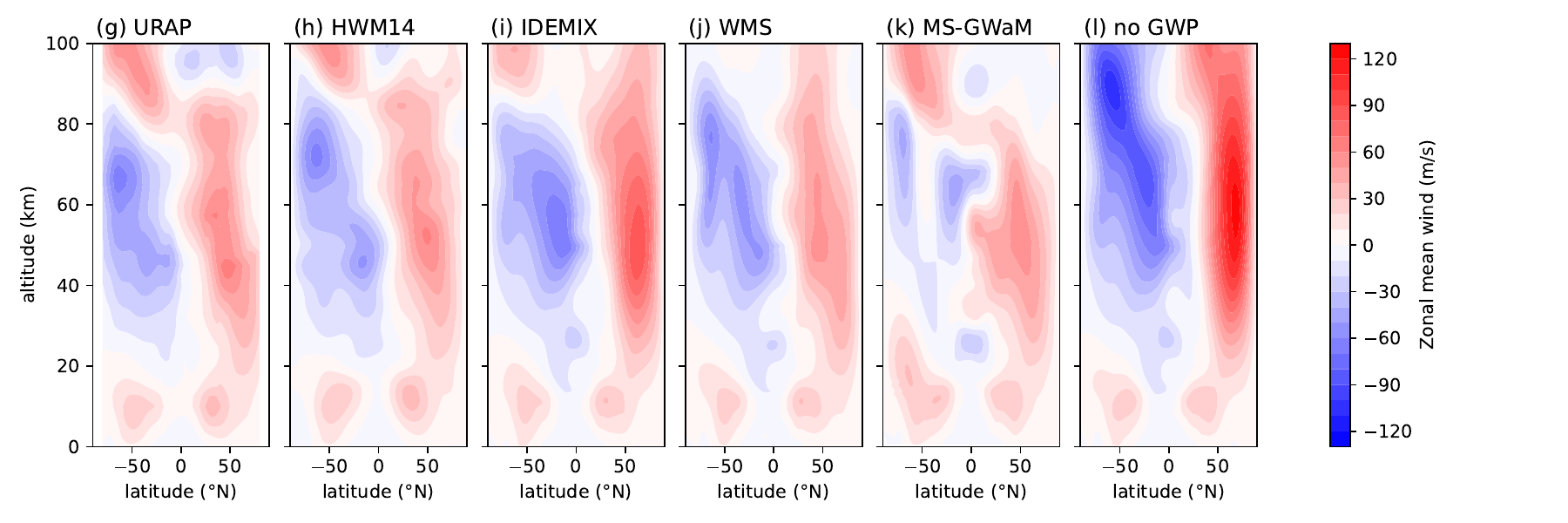}}

\caption{Zonal mean flow velocity (m/s) top: June 1991 and bottom: December 1991 for smaller IDEMIXforcing flux $F_{bot} = 2\times 10^{-3}$Jm$^{-2}$s$^{-1}$, $K_0 = 2\times 10^{-4}$m$^{-1}$s$^{-1}$, $m_s = 7\times 10^{-4}$m$^{-1}$ and $m_* =2\pi/10$km. }
\label{U_jun_dec_fm2e-3}
\end{figure}

\begin{figure}
\centerline{
\includegraphics[width = 0.8\columnwidth]{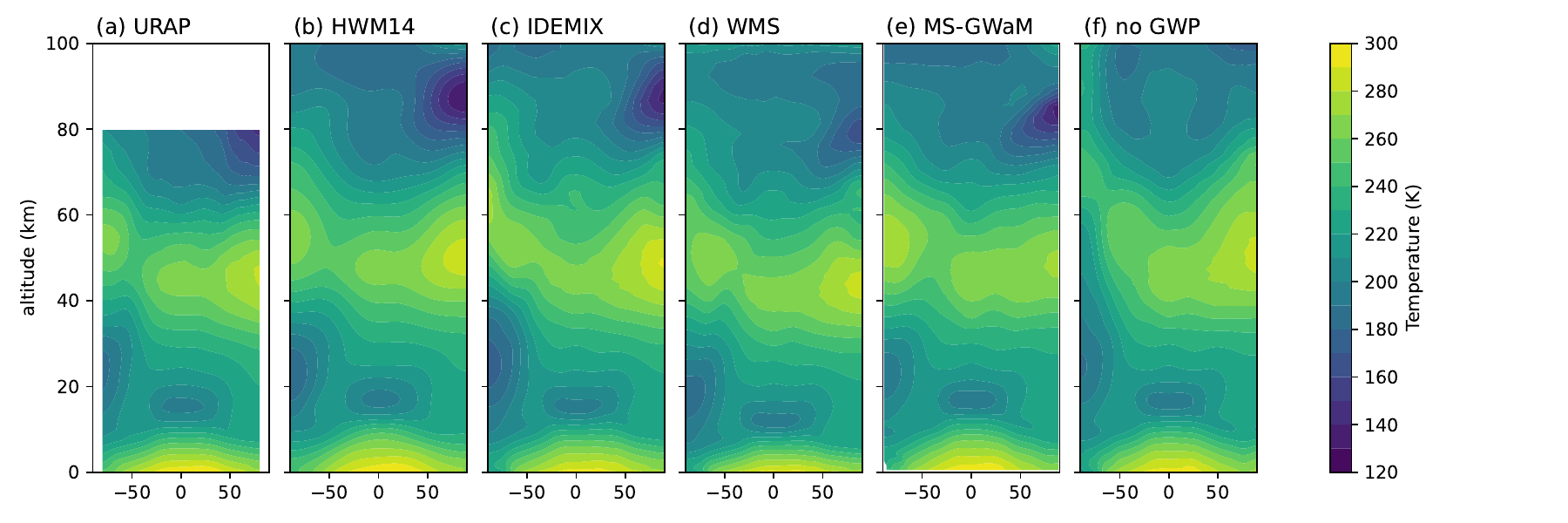}}

\centerline{\includegraphics[width = 0.8\columnwidth]{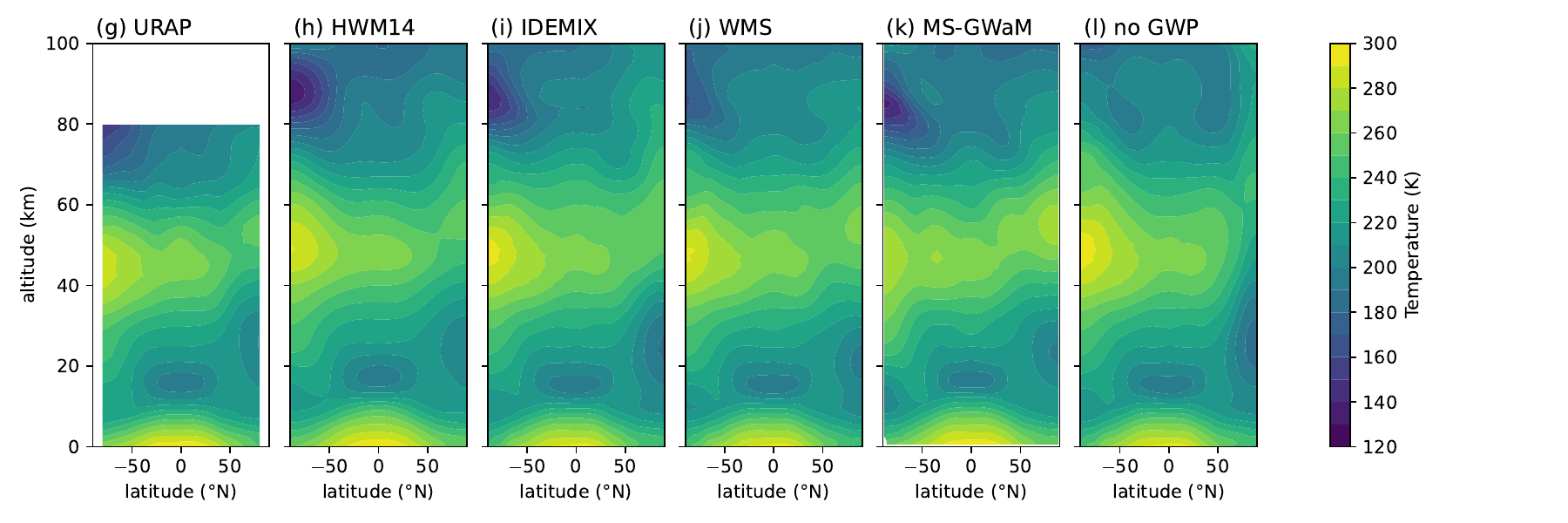}}

\caption{ Temperature (K) in top June, bottom December 1991. Forcing flux $F_{bot} = 2\times 10^{-3}$Jm$^{-2}$s$^{-1}$, $K_0 = 2\times 10^{-4}$m$^2$s$^{-1}$, $m_s = 7\times 10^{-4}$m$^{-1}$ and $m_* =2\pi/10$km.}
\label{temp_jun_dec_fm2e-3}
\end{figure}

Fig.~\ref{U_jun_dec_fm2e-3}(i) shows that using wave forcing equivalent to $0.025mPa$ in IDEMIX, in December there is improved closure of the winter jet around 100km.  However, the strength of the jet is a little too strong due to the waves travelling higher into the atmosphere before they break, and subsequently less GWD in the stratosphere and lower mesosphere.  The mean flow reversal in the summer hemisphere also occurs at a better altitude, resulting in the summer cold pole at an altitude more agreeable to the HWM14 data, as shown in Figs.~\ref{temp_jun_dec_fm2e-3}(c) and (i).  However it does mean losing the strong stratopause winter polar maximum temperature.

\begin{figure}
\includegraphics[width =0.5\columnwidth]{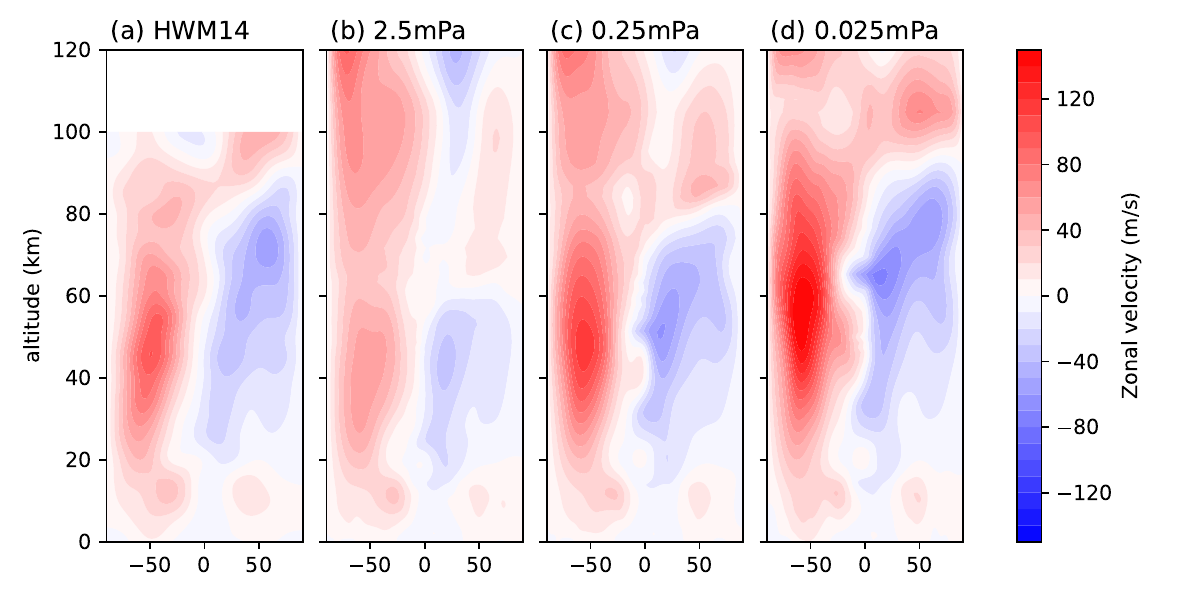}
\includegraphics[width =0.5\columnwidth]{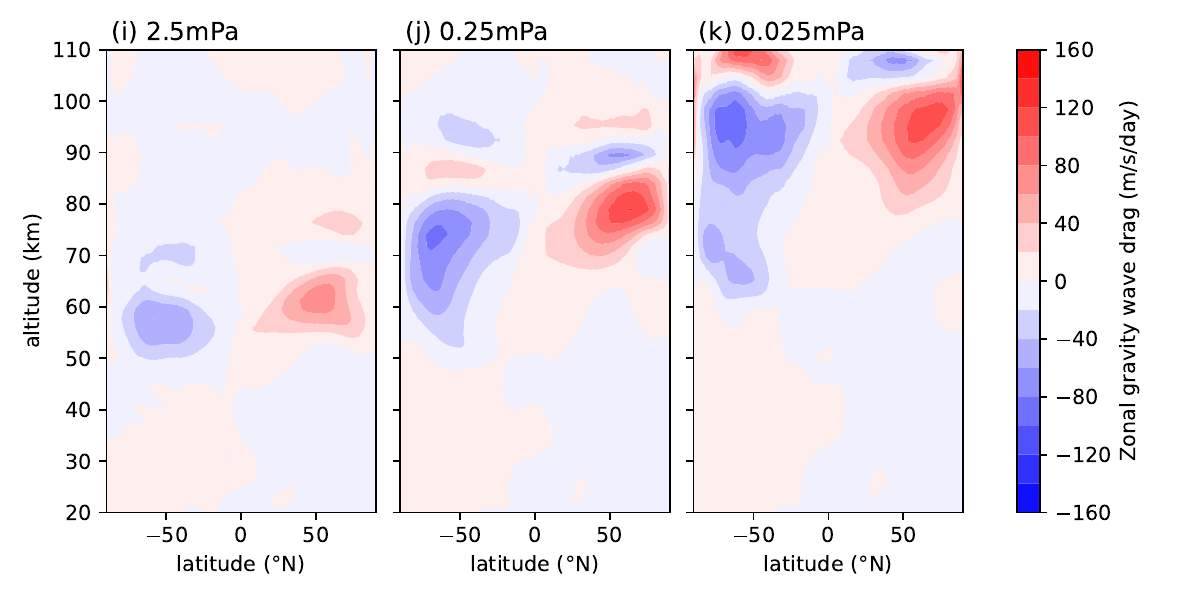}\\
\includegraphics[width =0.5\columnwidth]{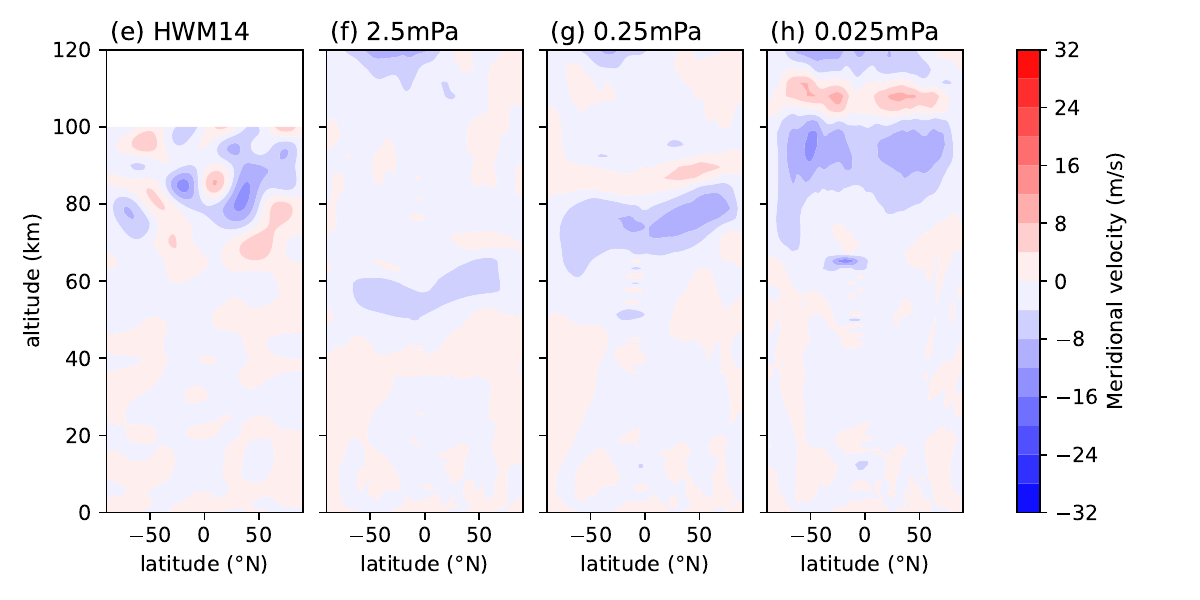}
\includegraphics[width = 0.5\columnwidth]{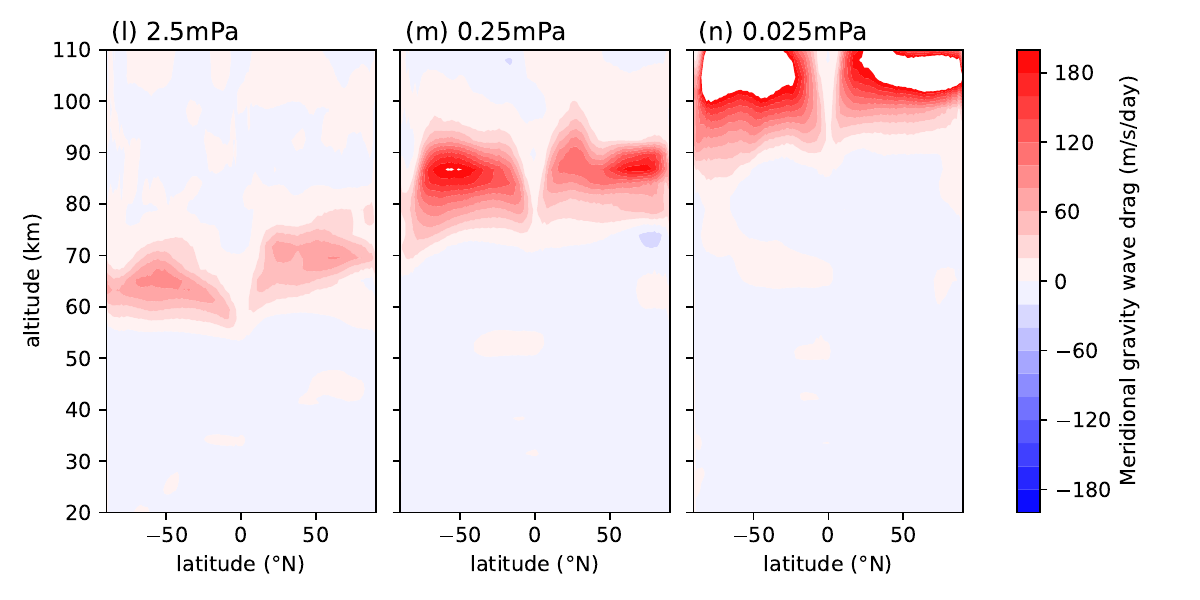}
\caption{ Top: Zonal mean flow velocity (m/s) and zonal gravity wave drag(m/s/day) bottom: meridional flow velocity (m/s) and meridional GWD (m/s/day) for June 1991 for various levels of IDEMIX forcing. $K_0 = 20$m$^2$s$^{-1}$, $m_s = 4.5\times 10^{-4}$m$^{-1}$ and $m_* =2\pi/15$km.}
\label{UV_jun_fmcomp}
\end{figure}

All of these traits are due to the GWs travelling higher into the atmosphere before reaching saturation point.   There is less GWD at lower altitudes to dampen the main jet and more GWD available at higher altitudes to close the jets.  In the winter hemisphere, for both June and December, the negative GWD is reduced and leads to a small slice of positive GWD at about 105km.  This effect is more clearly clarified in Fig.~\ref{UV_jun_fmcomp} for various forcing levels for June.  With 2.5mPa forcing, the GWD, both zonal and meridional are centred around 60km altitude.  But the corresponding mean flow plots show that this leads to the mean zonal flow reversal too low in the middle atmosphere.  Also the meridional mean flow adopts a more layered structure for the successively smaller forcing. While the layered structures do not closely resemble the HWM14 data so much, the altitude at which these structures occur at least better match the data with decreased forcing.  For IDEMIX, the ideal forcing is
between 0.25mPa and 0.025mPa, to provide sufficient momentum flux both in the middle atmosphere and in the lower thermosphere.

%

\subsection{Effects of $m_*$}

The standard spectral bandwidth parameter, especially for the launch spectrum, often cited in the literature is $m_*=2\pi/2$km, adopted from sparse observations.  

\begin{figure}
\includegraphics[width =\columnwidth]{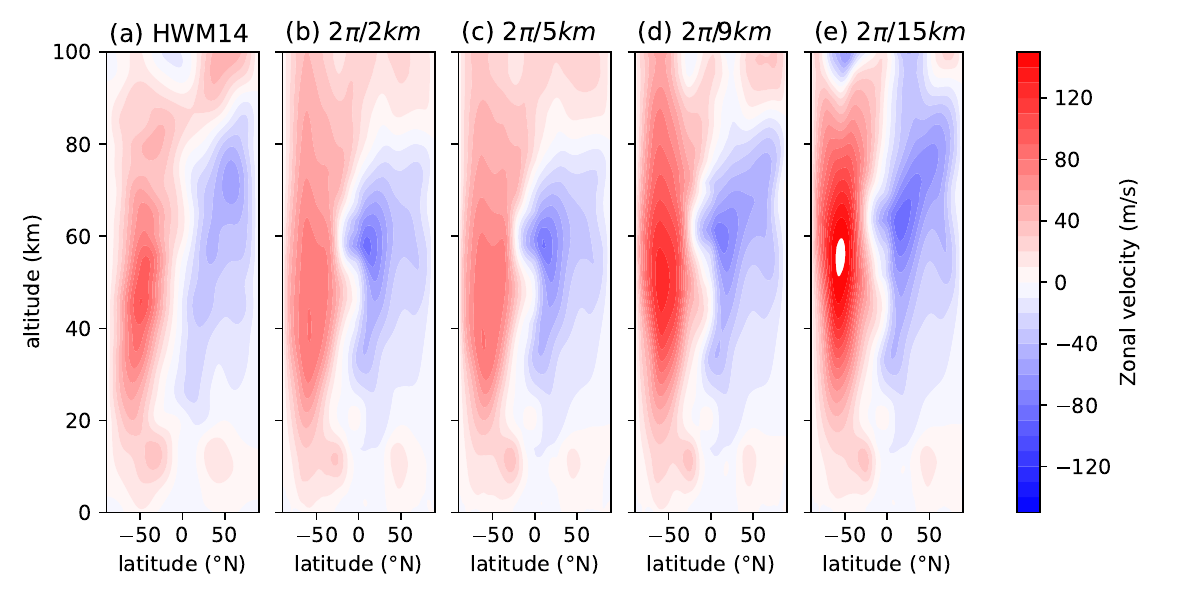}
\caption{ 
Zonal mean flow velocity (m/s) and zonal gravity wave drag(m/s/day) 
for June 1991.  Forcing flux $F_{bot} = 2\times 10^{-3}$Jm$^{-2}$s$^{-1}$, $K_0 = 2\times 10^{-4}$m$^2$s$^{-1}$ and for various levels of bandwidth parameter $m_* =2\pi/2$km, $2\pi/5$km, $2\pi/9$km and $2\pi/15$km.}
\label{UV_jun_mstcomp}
\end{figure}

For IDEMIX, Figs.~\ref{UV_jun_mstcomp}(b) and (c) show that the larger values of $m_*=2\pi/2$km and $2\pi/5$km give better results in the stratosphere, where the saturated waves provide the GWD to damp the intensity of the winter jet to agree better with the HWM14 data in Fig.~\ref{UV_jun_mstcomp}(a).  Since the GWD deposition is at lower altitudes, the reversal the summer jet is also lower than it should be.  In the mesosphere however, the summer jet has an improved tilt and the zonal mean wind reversal has a more realistic altitude for the smaller values, such as $m_*=2\pi/9$km as evident in Fig.~\ref{UV_jun_mstcomp}(d).
The smaller $m_*$ also improves the closure of the winter jet above the mesopause.  These results support the early suggestions~\cite{VanZandt1982,FrittsVanZandt1993} that $m_*\approx2\pi/2$km around the tropopause and $m_*\approx2\pi/20$km around the mesopause.

\section{Summary and Discussion}

A novel transient gravity wave parameterization, IDEMIX, which is derived from the wave energy balance equation, has been developed for the atmosphere and implemented into a general circulation model.  This  addresses the need for improved sophistication of GWPs.
It has been shown that IDEMIX can replicate well the mean circulation in the middle atmosphere and lower thermosphere, exhibiting mean flow reversals and the cold summer mesopause.  

The energetically-balanced model includes the vertical propagation of the wave field, it's interaction with the background velocity field, a critical layer closure and wave dissipation when the wave field becomes saturated.  A fixed spectrum is assumed and all wavenumbers and frequencies are accounted for.  Results shown here adopt the Desaubies spectrum and support the idea that the spectral bandwidth is larger at the tropopause and smaller at the mesopause.  The increased physical processes in this parameterization do not amount to a higher numerical cost as IDEMIX is 35\% faster than the WMS scheme, when both schemes have four azimuthal compartments.  The one-dimensional wave ray-tracing MS-GWaM is about five times slower than WMS.

Noteworthy with IDEMIX, is the reversal of the zonal GWD around the mesopause region.  This GWD reversal, not simulated by WMS or MS-GWaM, is observed in the high resolution GW-resolving simulations presented by~\citeA{BeckerVadas2018}, where they hypothesise that this is due to secondary gravity wave emission after primary wave breakdown.  It is also observed in the Middle and Upper Atmosphere Model~\cite{Lilienthaletal2020} with a nonlinear spectral whole atmosphere GWP~\cite{Yigitetal2008} for January conditions where it is suggested that the faster GWs are necessary to influence circulation patterns and temperature differences in the lower thermosphere, such as the layered meridional wind structures, observed here by IDEMIX also for smaller forcing magnitude.  Here, it is suggested that this GWD reversal will only be present when there is an increase in wave momentum flux at these higher altitudes, whether that comes from the further-travelling high-phase speed, smaller amplitude primary waves' interaction with the mean circulation or from secondary gravity wave sources.

IDEMIX has the potential to be further developed to include lateral wave propagation as well as multiple wave sources.


\appendix
\section{IDEMIX coefficients}
\label{apx_coeffs}

\label{IDEMIXcoeffs}
Denoting $x=N/f$, the integrated mean group velocity is,
\begin{eqnarray}
\label{c0incode}
c_0 = \sqrt{2} \frac{f}{m_*} \frac{12 x^{2/3} }{35 (x^2-1)(x^{2/3}-1)}
\left( 5 (x^{7/3}- 1) + 7 x^{1/3} (1- x^{5/3}) \right)
\end{eqnarray} 
and the mean flow interaction coefficient is
\begin{eqnarray}
\Lambda  = 
\end{eqnarray}

From the CL parameterisation in Eq.~\eqref{alphacl}, and writing $\alpha = m_s A(-m_s) \Phi$, then
\begin{eqnarray} \nonumber
\Phi&=& \frac {9\sqrt [3]{2}}{4\sqrt [3]{x}\pi\, \left( {x}^{2/3}-1
 \right)  \left( \Gamma \left( 2/3 \right)  \right) ^{3}} \nonumber
 \\ && \nonumber
\left( 
{\frac {8}{81}\sqrt {3}\sqrt [3]{2}{\pi}^{3}
\sqrt [3]{x}{\mbox{$_2$F$_1$}(-1/2,1/3;\,5/6;\,{x}^{-2})}}
-{\mbox{$_2$F$_1$}(-1/3,1/2;\,7/6;\,{x}^{-2})} \left( \Gamma \left( 2/3
 \right)  \right) ^{6}
 \right)
 \end{eqnarray}

  \begin{acronyms}
  \acro{IDEMIX}
   Internal wave Dissipation, Energy and Mixing
  \acro{GWP}
  Gravity wave parameterization
  \acro{MS-GWaM}
  Multiscale gravity wave model
  \acro{ICON}
  Icosahedral nonhydrostatic
  \acro{UA-ICON}
  Upper atmosphere icosahedral nonhydrostatic
  \acro{ECMWF}
  European Centre for Medium-Range Weather Forecasts
  \acro{GWD}
  Gravity wave drag
  \acro{WMS}
  Warner McIntyre Scinocca
  \acro{UARS}
  Upper Atmosphere Research Satellite
  \acro{URAP}
  UARS Reference Atmosphere Project
  \acro{HWM}
  Horizontal wind model
  \end{acronyms}

%
%

\section*{Open Research Section}

The ICON software is freely available to the scientific community for noncommercial research purposes under a license from DWD and MPI-M. Potential users who would like to obtain ICON can contact icon@dwd.de. 

The IDEMIX code and its module for implementing into ICON have been developed at the University of Hamburg, Germany.  It is available on the Zenodo repository~\cite{QuinnVoelker2023}, as well as the python scripts for generating the figures.
The URAP wind and temperature data are available online (https://www.sparc-climate.org/data-centre/data-access/reference-climatology/urap/), as are the HWM14 (https://ccmc.gsfc.nasa.gov/models/HWM14~2014/).



\acknowledgments
This paper is a contribution to the
subproject W1 (Gravity wave parameterization for the atmosphere) of the 
Collaborative Research Centre TRR 181 ``Energy Transfer in Atmosphere and Ocean" funded by the Deutsche Forschungsgemeinschaft (DFG; German Research Foundation)—Project 274762653.  Ulrich Achatz thanks the DFG for partial support through CRC 301 ``TPChange" (Project-ID 428312742, Projects B06 “Impact of small-scale dynamics on UTLS transport and mixing” and B07 “Impact of cirrus clouds on tropopause structure”).


\bibliography{main}

\end{document}